\documentclass[aps,prb,reprint]{revtex4-1}
\usepackage{amsmath}
\usepackage{units}
\usepackage{url}
\usepackage{amsfonts}
\usepackage{amssymb}
\usepackage{textcase}
\usepackage{bm}
\usepackage{natbib}
\usepackage{graphicx}
\usepackage{braket}
\usepackage{color}
\newcommand{\pvec}[1]{\vec{#1}\mkern2mu\vphantom{#1}}
\renewcommand{\vec}[1]{\boldsymbol{#1}}

\begin{document}
\title{Metastable electron--electron states in double--layer graphene structures}
\author{L.L. Marnham}
\email{llm204@exeter.ac.uk}
\affiliation{School of Physics, University of Exeter, Stocker Road, Exeter EX4 4QL, United Kingdom}
\author{A.V. Shytov}
\affiliation{School of Physics, University of Exeter, Stocker Road, Exeter EX4 4QL, United Kingdom}
\pacs{81.05.U-,71.10.Li,31.15.ac,71.35.-y}

\begin{abstract}
The prototypical exciton model of two interacting Dirac particles in graphene was analyzed in Ref.~[\onlinecite{sabio10}]
and it was found that in one of the electron--hole scattering channels the total kinetic energy vanishes, 
resulting in a singular behaviour. We show that this singularity can be removed by extending the quasiparticle dispersion, 
thus breaking the symmetry between upper and lower Dirac cones. The dynamics of an electron--electron pair are then mapped 
onto that of a single particle with negative mass and anisotropic dispersion. We show that the interplay between dispersion 
and repulsive interaction can result in the formation of bound, Cooper--pair--like, metastable states in double--layered hybrid structures. 
\end{abstract}
\maketitle

\section{Introduction}
Graphene continues to receive significant attention for its numerous intriguing interaction effects and transport properties~\cite{katsnelson06,shytov07,gu11,beenakker06}. Recent experiments with high--quality  samples suggest the existence of non--trivial correlated phases in graphene, such as excitonic condensates in non--zero magnetic fields~\cite{gorbachev12}. However, the zero--field condensate predicted theoretically \cite{min08,zhang08,abergel2013} has not been observed~\cite{gorbachev12,kim11}. This has provoked significant interest in the archetypical two--body problem~\cite{sabio10,lee12,mahmoodian13}.

Most interesting properties of graphene are due to the existence of Dirac cones~\cite{wallace47} 
located near two points in the Brillouin zone, $\vec{K}^+$ and $\vec{K}^-$. Each cone hosts positive and negative 
energy states, each with linear dispersion $\epsilon=\pm v_Fp$, akin to electron and positron states 
in quantum electrodynamics. (Here $v_F=10^6ms^{-1}$ is the Fermi velocity\cite{deacon07} and $p$ is the magnitude 
of the momentum $\vec{p}=(p_x,p_y)$.) The symmetry between positive and negative cones results in the compensation 
of total kinetic energy for two particles with opposite momenta. Thus, the two--particle states can be divided into 
\textit{dispersing} (spanned by states where $E\neq 0$) and \textit{non--dispersing} ($E=0$) sectors. 
States in the non--dispersing sector have momentum--independent eigenvalues, and are therefore infinitely degenerate. 
However the linear dispersion is only accurate at low energies; higher order terms can at times reveal important physics hidden by the conical approximation.

The effect of dispersion on the excitonic physics can be seen if one considers Dirac particles interacting via the Coulomb potential,
which scales as $U(r)\sim\frac{Ze^2}{\epsilon r}$. In the conical approximation, the kinetic energy scales as $v_Fp\sim \frac{\hbar v_F}{r}$ due to the uncertainty principle. 
In the case of a single particle interacting with a static charge, Dirac vacuum reconstruction occurs when the potential energy dominates 
over kinetic energy\cite{shytov07}: $\frac{Ze^2}{\epsilon}>\frac{\hbar v_F}{2}$. In the case of two carriers, however, a doubling of the 
relative velocity effectively doubles the critical charge for which collapse is possible\cite{sabio10}: $Z_C\gtrsim 2$. This renders 
the strongly--interacting regime irrelevant for electron--hole physics. For a weaker, quadratic dispersion ($p^2\sim\frac{\hbar^2}{r^2}$), 
however, bound states can form for arbitrarily small interaction strength. For this reason, the previously neglected contribution of non--dispersing states \cite{sabio10,lee12} can be 
important for understanding the two--particle physics of graphene. To this end, the two--body problem 
was re--analyzed in Ref.~[\onlinecite{mahmoodian13}] with the inclusion of trigonal--warping terms which preserve the symmetry between the two cones 
but do not lead to non--zero kinetic energy when the total momentum of the pair is zero. In this paper we introduce quadratic momentum terms due to next--nearest--neighbour hopping which were ignored by Ref.~[\onlinecite{mahmoodian13}] 
and show that this leads to the finite kinetic energy necessary for bound state formation. In particular, we will show that this leads to a new class of states which exist regardless of the orientation in momentum space (and cannot arise due to trigonal terms alone due to sign--indefinite kinetic energy). The formation of pairs of particles in the same valley is allowed for the model we consider here, which was not the case for the electron--hole case in Ref.~[\onlinecite{mahmoodian13}].\\
\indent In this work, we show that Cooper--pair--like states can be formed in the subspace of non--dispersing two--particle states. The dynamics in this sector is governed by quadratic terms in the single--particle dispersion. Two such contributions are possible: an isotropic term due to next--nearest--neighbour hopping, $\epsilon_I\propto p^2$, and an anisotropic term due to trigonal--warping, $\epsilon_A\propto p^2\sin(3\phi_{\vec{p}})$, where $\phi_{\vec{p}}$ is the polar angle in momentum space defined by $\tan(\phi_{\vec{p}})=\frac{p_y}{p_x}$. We show that, depending upon the relative magnitudes of these two terms, two regimes are possible. When the isotropic contribution dominates, bound states can be formed; otherwise it is possible to form non--dispersing quasibound states (which can leak into the continuum.) We calculate the binding energies of such states numerically, for a double--layer configuration, and discuss the decay rate due to coupling to the continuum of dispersing states.\\
\indent The rest of this work shall be structured as follows. In Sec. II we construct the effective Hamiltonian of a pair of interacting electrons in double--layer graphene. In particular, we discuss the inclusion of a finite band curvature into this Hamiltonian, before projecting out the high energy states and focusing on the non--dispersing sector discussed above. In Sec. III we will calculate the binding energies of these pairs, approaching the problem from two directions. Firstly, we approximate the energies by treating the potential within the harmonic approximation. Secondly, we find the direct--pair energies by calculating the local density of states numerically. We find that these approaches give an order--of--magnitude agreement, with our numerical method yielding $E=45$meV for the bound state of highest energy. In Sec. IV we present an analysis of the semiclassical trajectories of the pair, which gives an intuitive view of how a pair can become bound in the presence of a repulsive interaction. In Sec. V we discuss the coupling of these states to the dispersing sector. Although the potential can destroy the states in principle, the decay rate vanishes by symmetry for the highest energy level of the pair. The kinetic energy also leads to a decay, but we argue that the decay rate is small enough to validate our consideration of the non--dispersing sector in isolation. Finally, in Sec. VI we summarize our results. Discussions of the interlayer electron--electron interaction and our numerical approach to calculating the local density of states can be found in the appendices.
\section{Effective Hamiltonian}
\indent We begin by analyzing the kinetic energy of two Dirac quasiparticles in graphene. Since electrons in graphene can reside on two sublattices, $A$ and $B$, they are to be described by a two--component Dirac spinor. The internal degree of freedom arising due to the presence of the sublattices is known as pseudospin (for a discussion, see Ref. [\onlinecite{katsnelson12}]). In the low--energy approximation, the Dirac spinors for the two valleys, $\vec{K}^+$ and $\vec{K}^-$, can be treated as fully independent. We define these spinors as $\psi_{\vec{K}^+}=\left[\psi_{\vec{K}^+}^A,\psi_{\vec{K}^+}^B\right]^T$ and $\psi_{\vec{K}^-}=\left[\psi_{\vec{K}^-}^B,\psi_{\vec{K}^-}^A\right]^T$, where $A$ and $B$ label the probability amplitudes for the two sublattices. In the conical approximation, the dynamics of the pair is governed by the dispersion arising from the relative motion of its constituent particles, $\widehat{H_L}=v_F\vec{\sigma}_1\cdot\hat{\boldsymbol{p}}_1+v_F\vec{\sigma}_2\cdot\hat{\boldsymbol{p}}_2$, where $\boldsymbol{\sigma}_i$ is the pseudospin operator, subscripts denote the particle number and $\hat{\boldsymbol{p}}_i$ is a small momentum measured with respect to the $\vec{K}^+$-- or $\vec{K}^-$--point. We focus on states with zero total momentum, such that $\vec{p}_1=-\vec{p}_2$. The eigenstates of $\widehat{H_L}$ are given by:
\begin{gather}\label{wf1}
\begin{aligned}
\ket{1,\phi_{\vec{p}}}&=\frac{1}{\sqrt{2}}\left[e^{-i\phi_{\vec{p}}}\ket{\uparrow\uparrow}+e^{i\phi_{\vec{p}}}\ket{\downarrow\downarrow}\right],\\
\ket{2,\phi_{\vec{p}}}&=\frac{1}{\sqrt{2}}\left[\ket{\uparrow\downarrow}+\ket{\downarrow\uparrow}\right],\\
\ket{3,\phi_{\vec{p}}}&=\frac{1}{2}e^{i\phi_{\vec{p}}}\ket{\downarrow\downarrow}-\frac{1}{2}e^{-i\phi_{\vec{p}}}\ket{\uparrow\uparrow}+\frac{1}{2}\left[\ket{\uparrow\downarrow}-\ket{\downarrow\uparrow}\right],\\
\ket{4,\phi_{\vec{p}}}&=\frac{1}{2}e^{-i\phi_{\vec{p}}}\ket{\uparrow\uparrow}-\frac{1}{2}e^{i\phi_{\vec{p}}}\ket{\downarrow\downarrow}+\frac{1}{2}\left[\ket{\uparrow\downarrow}-\ket{\downarrow\uparrow}\right].
\end{aligned}\raisetag{4\baselineskip}
\end{gather}
\noindent The vectors $\ket{\uparrow}$ and $\ket{\downarrow}$ here represent the up-- and down-- pseudospin states for a single particle, so that the two--particle eigenstates are, e.g., $\ket{\uparrow\downarrow}=\ket{\uparrow}\bigotimes\ket{\downarrow}$. The states in Eq. (\ref{wf1}) have corresponding eigenvalues $E_{1,2}=0$ and $E_{3,4}=\pm 2v_Fp$. The subspace spanned by $\ket{1,\phi_{\vec{p}}}$ and $\ket{2,\phi_{\vec{p}}}$ from Eq. (\ref{wf1}) is the \textit{non--dispersing sector}. Such states are formed by electron quasiparticles in opposite cones, with the same magnitude of momentum, so that the relative velocity of the pair vanishes. Similarly, the subspace spanned by $\ket{3,\phi_{\vec{p}}}$ and $\ket{4,\phi_{\vec{p}}}$ forms the \textit{dispersing sector}, in which the velocities are opposite. In the absence of interactions, all states in the non--dispersing sector are infinitely degenerate.\\
\indent We note that this degeneracy is lifted if the symmetry between the upper and lower cones is broken, e.g., by a small band curvature. We extend the kinetic energy by quadratic terms compatible with the symmetries of the honeycomb lattice (see Refs.~[\onlinecite{katsnelson12}] and~[\onlinecite{,mccann06c}]). We write the single--particle kinetic energy in the form:
\begin{gather}
\begin{aligned}
\widehat{H}_{j}&=v_F\vec{\sigma}_j\cdot\vec{p}_j-\frac{p^2_j}{4m^*}+\tau_j\mu (p_{x,j}+ip_{y,j})^2\sigma_{+,j}+\text{H.c.},\label{single}
\end{aligned}\raisetag{0.5\baselineskip}
\end{gather}
\noindent where $j$ is the particle number, $p_j^2=p_{x,j}^2+p_{y,j}^2$,  $\sigma_{+,j}=\frac{1}{2}\left(\sigma_{x,j}+ i\sigma_{y,j}\right)$, $\tau_j=\pm 1$ for an electron in the $\vec{K}^{\pm}$ valley (determining the sign of the trigonal--warping) and H.c. denotes the Hermitian conjugate. The second term in Eq. (\ref{single}) is invariant under all two--dimensional rotations, and arises microscopically from contributions due to the hopping of electrons from one atom to its next--nearest--neighbour, giving $m^*=\frac{\hbar^2}{9a^2t'}$, where $t'$ is the next--nearest--neighbour hopping parameter \cite{castro09}.  The third term (including H.c.) is invariant under rotations by $120^\circ$. This term represents trigonal--warping, and originates from nearest--neighbour hopping, expanded to second order in momentum\cite{castro09,katsnelson12}, so that $\mu=\frac{3a^2t}{8\hbar^2}$.\\
\indent To examine the dynamics in the non--dispersing sector, we restrict the two--particle Hamiltonian to this subspace. We explicitly treat two distinct cases: \textit{direct pairs} (when both particles are in the same valley) and \textit{indirect pairs} (opposite valleys). All states $\ket{1,\phi_{\vec{p}}}$ and $\ket{2,\phi_{\vec{p}}}$ are annihilated by the operator $(\vec{\sigma}_1-\vec{\sigma}_2)\cdot\vec{p}$. Calculating the matrix elements of the kinetic energy we find the effective Hamiltonian:
\begin{align}
\widehat{H}_{1,2}^\text{eff}=\left[\begin{array}{cc}
-\frac{p^2}{2m^*} & \tau_{1,2}\mu p^2\sin(3\phi_{\vec{p}}) \\
\tau_{1,2}\mu p^2\sin(3\phi_{\vec{p}})  & -\frac{p^2}{2m^*}
\end{array}\right]\label{direct},
\end{align}
\noindent where the rows and columns correspond to states $\ket{1,\phi_{\vec{p}}}$ and $\ket{2,\phi_{\vec{p}}}$ and $\tau_{1,2}=\tau_1+\tau_2$.\\
\indent We will show that some of the features in the dynamics of two--particle states crucially depend upon the signs and relative magnitudes of the quadratic terms, i.e., on the values of $m^*$ and $\mu$. It has been shown by a variety of different approaches \cite{deacon07,castro09,reich02,kretinin13,kuhne12} that $t$ and $t'$ have the same sign, however there is a disagreement on the precise value of $t'$. \textit{Ab initio} calculations\cite{castro09, reich02} give the range $0.02t\leq t'\leq 0.2t$, while cyclotron resonance\cite{deacon07}, quantum capacitance \cite{kretinin13} and polarization--resolved magnetospectroscopy \cite{kuhne12} measurements have produced $t'=0.04t,0.11t\text{ and }0.14t$ respectively. The full two--particle kinetic energy is $\widehat{H}_{1,2}=\widehat{H}_{1}+\widehat{H}_{2}$, and so $-m^*$ plays the role of a two--particle reduced mass due to the $-\frac{p^2}{2m^*}$ term which arises when $\widehat{H}_{1,2}$ is written explicitly, with the corresponding range of values $0.7\leq\frac{m^*}{m_e}\leq 7.5$. This implies that the isotropic kinetic energy term is negative definite, which will be shown to be of crucial importance to the spectrum of two--particle states.\\
\section{Bound states of electron pairs}
\begin{figure}[t]
\includegraphics[width=\linewidth]{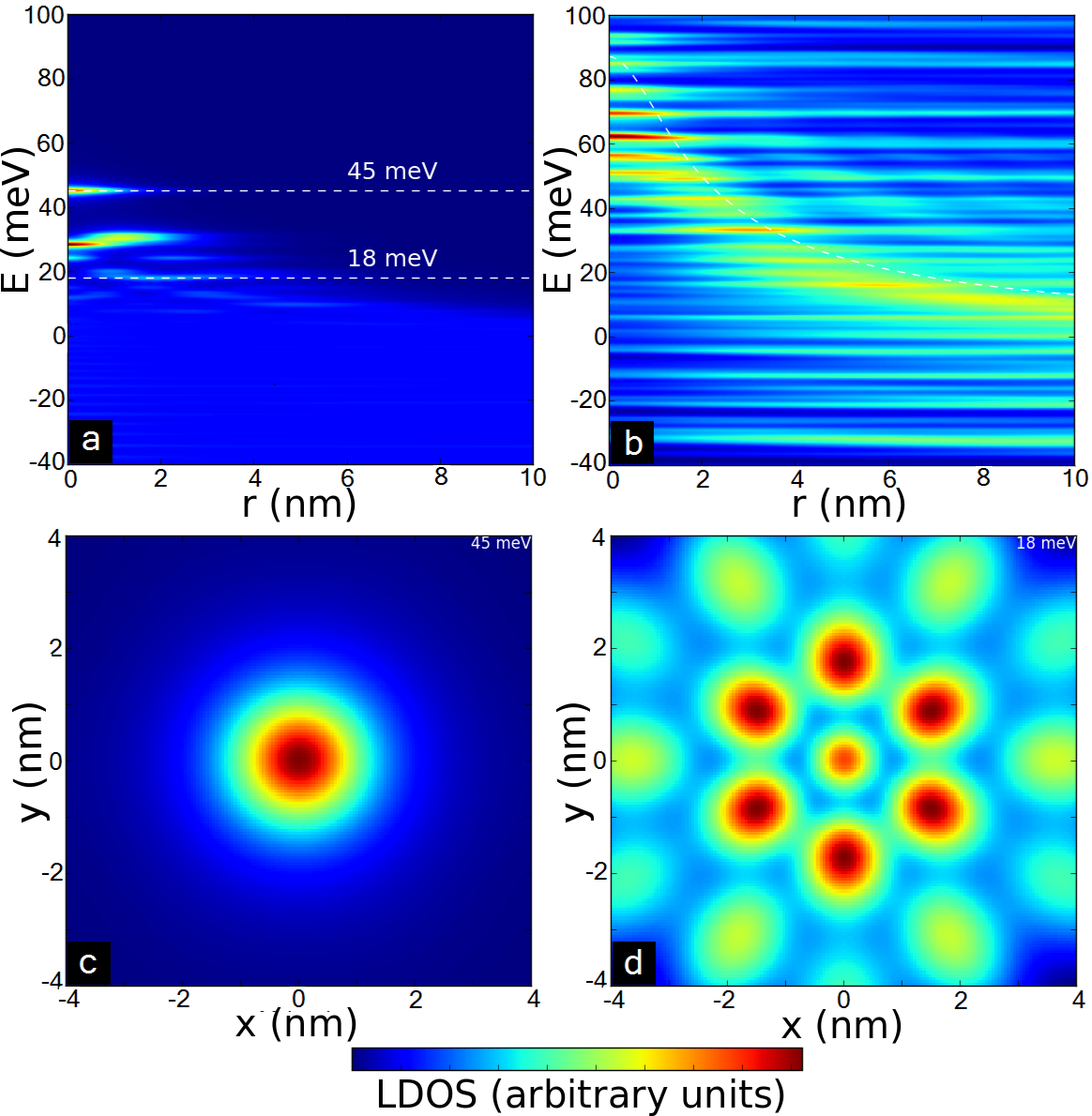}
\caption{Two--body LDOS in a graphene hybrid structure as a function of binding energy $E$ and interparticle distance $r$. The energy dependence shows distinct bound states in the (a) isotropic regime, which peel off into the free particle continuum in the (b) anisotropic regime (Coulomb potential in white). In real space, the wavefunctions are those of highly localized bound states at (c) $\epsilon_1=45$meV, and have higher anisotropy for (d) less bound states.}
\label{LDOS_fig}
\end{figure}

\indent To understand the dynamics of pairs described by the kinetic energy terms in $\widehat{H}_{1,2}$, let us first consider the simplest case of indirect pairs, where the electrons are in opposite valleys. In this configuration $\tau_{1,2}=0$, so that the contribution of trigonal--warping vanishes and the only remaining kinetic term is $-\frac{p^2}{2m^*}$. The dynamics of the interacting pair is therefore described by the Hamiltonian $H_I=-\frac{p^2}{2m^*}+U(r)$ for states with configuration $\ket{2,\phi_{\vec{p}}}=\frac{1}{\sqrt{2}}[\ket{\uparrow\downarrow}+\ket{\downarrow\uparrow}]$, where $U(r)$ is the potential energy. The Hamiltonian $H_I$ describes the motion of a particle with \textit{negative} effective mass $-m^*$ in the external potential $U(r)$. We note that $-H_I$ describes the motion of a particle with a \textit{positive} mass $m^*$ in an attractive potential. In two dimensions, an arbitrarily weak attractive potential exhibits at least one bound state at negative energies for massive particles\cite{landau77}. It follows, therefore, that $H_I$ will exhibit positive energy bound states if $U(r)$ is repulsive. This property is a direct consequence of the negative definite kinetic energy of the pair, the dynamics of which is akin to the motion of a hole--like state near the top of the valence band in a semiconductor: the repulsive potential due to a negatively charged impurity is perceived as an attraction due to negative band curvature. In the real space picture, two electrons with opposite momenta reside in different cones and have nearly the same velocities. The repulsive force tends to increase the momentum of one electron, and decrease the momentum of the other. Due to the negative dispersion term, this decreases the velocity of the first particle, and increases the velocity of the other, reducing the distance between them. Unlike conventional bound states, these positive eigenstates are metastable. Formation of regular (electron--hole) excitons is prohibited in this regime.\\
\indent Metastable states similar to the ones described by $H_I$ were previously discovered in connection with inverse hydrogen absoption spectra \cite{gross71}. Simple models of bound state formation due to a negative single--particle energy dispersion near the top band boundary were considered in Refs.~[\onlinecite{mahajan06}] and~[\onlinecite{souza10}]. We note, however, that the origin of the negative dispersion in graphene is different: the leading term in the single particle energy is linear, and the Hamiltonian $H_I$ arises via the compensation between the two sub--bands. This means that the dynamics of these pairs can be represented as a slow relative motion $v\propto p/m^*$ equation superimposed with the fast motion of the pair, $v\propto v_F$.\\
\indent Since the effective mass ($-m^*$) is only about five times larger than the free electron mass, the binding can be quite strong. For the example of a repulsive Coulomb interaction, $U(r)=\frac{e^2}{\epsilon_s\epsilon_N r}$, the problem reduces to the two--dimensional hydrogen atom\cite{zaslow67}. (Here $\epsilon_N=1+\frac{N\pi e^2}{8\epsilon_s\hbar v_F}$ is the intrinsic dielectric constant\cite{hwang07} of graphene embedded in a material with dielectric constant $\epsilon_s$. For single (double) layer graphene, the number of fermion species is $N=4$ ($N=8$).) The highest energy level is given by
$E_1=\frac{2m^*e^4}{\epsilon_s^2\epsilon_4^2\hbar^2}$. We note that the hydrogen--like Hamiltonian $H_I$ results in binding energy $E_1\sim 1.5$eV and Bohr radius $a_B\sim 2.5\text{\AA}$ for $t'=0.1t$. At such short distances, the low--energy approximation to the graphene band structure is not valid \cite{geim07}, rendering the solution inconsistent. More importantly, the dynamics of particles at such high energies is affected by Pauli blocking due to the Dirac sea. For the bound state to be observable, the relevant phase space domain must be free from other particles. This can be achieved by, e.g., gating, if the bound state energy is well below $1$eV. Two--particle states with smaller binding energies can be realized in double--layered structures where the electrons in opposite layers are separated vertically by a dielectric spacer of thickness $d$. Hexagonal boron nitride ($\epsilon_{s}=3-4$) spacers have been experimentally shown to electrically isolate parallel graphene layers at a thickness of 4 atomic layers ($d=1.3$nm)\cite{britnell12}. This suppresses the $\frac{1}{r}$ singularity, yielding smaller binding energies. For a rough estimate of the binding energy we approximate the potential as $V(r)\sim e^2/\epsilon_s\epsilon_4^2\sqrt{r^2+d^2}$ (see Appendix A). The spectrum of the resulting shallow well can be found in the harmonic approximation, assuming $r\ll d$. The energy of the highest bound state is $E_0=-\hbar\omega+\frac{e^2}{\epsilon_s\epsilon_4^2d}$, where $\omega=\sqrt{\frac{e^2}{\epsilon_s\epsilon_4^2m^{*}d^3}}$ is the oscillation's angular frequency near the potential maximum. For example, $t'=0.1t$ gives binding energy $E_0=31$meV. For the case of direct pairs the trigonal warping is not compensated ($\tau_{1,2}\neq 0$), and the situation becomes more complicated. We note, however, that if the mass $m^*$ is small enough the trigonal warping terms cannot change the sign of the kinetic energy. Further, there are several momentum space orientations for which $\sin(3\phi_{\vec{p}})=0$. We therefore take $E_0$ as a first approximation of the binding energy for direct \textit{and} indirect pairs.\\
\indent To analyze the case of the direct pair with anisotropic dispersion, we derive its effective Hamiltonian in momentum space. The potential energy $V(r)$ is represented by a non--local operator proportional to its Fourier transform $\tilde{V}_{\vec{p},\pvec{p}'}=\tilde{V}(|\vec{p}-\pvec{p}'|)$, while the kinetic energy terms are given by Eq. (\ref{direct}) for $\tau_{1,2}=\pm 2$. Restricting the potential energy to the non--dispersing sector requires some care due to a non--trivial overlap between non--dispersing states with different momenta: $\braket{1,\phi_{\pvec{p}'}|1,\phi_{\vec{p}}}=\cos(\phi_{\vec{p}}-\phi_{\pvec{p}'})$. For the case of direct interactions, the $(\vec{p},\pvec{p}')$ block of the Hamiltonian matrix takes the form:
\begin{align}
\widehat{H}_{\vec{p},\pvec{p}'}=\delta_{\vec{p},\pvec{p}'}\widehat{H}_{1,2}^\text{eff}
+\tilde{V}_{\vec{p},\pvec{p}'}&\left[\begin{array}{cc}
\cos(\phi_{\vec{p}}-\phi_{\pvec{p}'}) & 0 \\
0 & 1\label{hamiltonian}
\end{array}\right],
\end{align}
\noindent where $\widehat{H}_{1,2}^\text{eff}$ is given by Eq. (\ref{direct}). In the absence of inter--particle interaction the eigenvalues of this matrix are given by the kinetic energy terms: $\epsilon_2^{(2)}=-2\mu p^2\left[\eta+\sin(3\phi_{\vec{p}})\right]$, where we have introduced the \textit{anisotropy parameter} $\eta=\frac{6t'}{t}$ which is not physically tunable (uncertainty in the value of $t'$ gives a range of possible values $0.12\leq\eta\leq 1.2$). Depending on the value of $\eta$, the kinetic energy is either negative--definite ($\eta>1$) or sign--indefinite ($\eta<1$). \\
\indent We proceed by numerically diagonalizing the Hamiltonian given by Eq. (\ref{hamiltonian}) using the interlayer interaction from Ref.~[\onlinecite{abergel2013}] (see Appendix A). Although it is assumed that the relevant phase space domain is free of other particles to avoid the effect of Pauli blocking, we will treat the case of screening at half--filling ($p_f=0$) as a first approximation. Indeed, the dielectric contribution to the screening giving rise to $\epsilon_N$ occurs at scales smaller than the Fermi wavelength ($\lambda_f$), and is most important in the realistic limit of $\lambda_f\ll d$. To visualize the resulting wavefunctions, we calculate the local density of states (LDOS), $\nu(\epsilon,x,y) =\sum_{n}\delta(\epsilon-\epsilon_{n})|\psi_{n}(x,y)|^2,$ where $x$ and $y$ are the components of the in--plane separation and $n$ labels the eigenstates. In the isotropic regime, $\eta>1$, as is evidenced by Fig. \ref{LDOS_fig}a for $\eta=1.1$, there is a formation of distinct, highly localized, bound states at $\epsilon_1=45$meV, $\epsilon_2=30$meV and $\epsilon_3=27$meV. We note that the energy of the highest bound state, $\epsilon_1$, is very similar to the value predicted in the harmonic approximation above. This validates the further use of such an approximation in the calculations of the transition rates that are to follow. At negative energies there is a low intensity continuum of unbound states, which are only weakly coupled to the bound states due to the symmetry of the Hamiltonian, resulting in their large lifetimes.\\
\indent In the anisotropic regime ($\eta< 1$), there are six \textit{easy axis} angles, defined by the relation $\sin(3\phi_{0})=-\eta$, along which the dispersion is effectively suppressed despite the broken conduction--valence symmetry. By concentrating the wavefunction along these axes, one constructs a state qualitatively similar to the non--dispersing solutions in which the interparticle distance takes a constant value $r_0$: $\psi(r)\propto\delta(r-r_0)$. The energies of these states (Fig. \ref{LDOS_fig}b) follow the profile of the interaction potential, $\epsilon\approx U(r_0)$. Further, a negative energy state dragged into the positive continuum by the interaction potential can decay by changing its pseudospin configuration rather than by tunnelling through a barrier. This is wholly due to the sign--indefinite kinetic energy.\\
\section{Semiclassical trajectories}
\begin{figure}[h]
\includegraphics[scale=0.35]{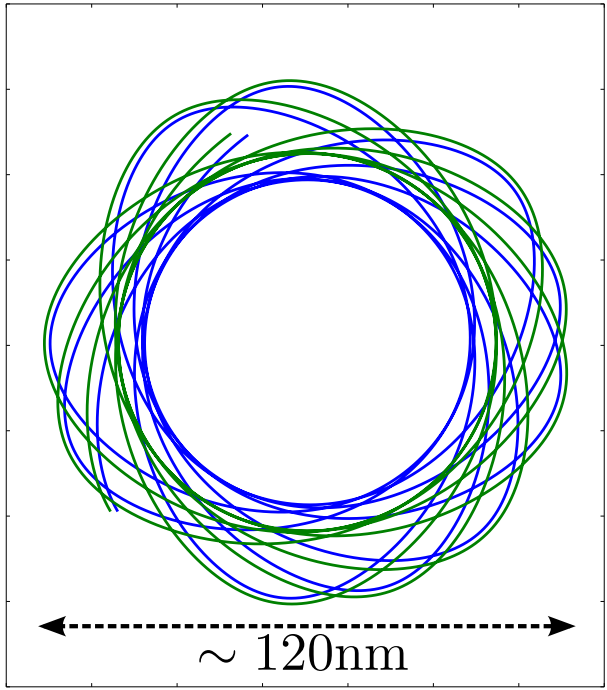}
\centering
\caption{(Color online) Typical trajectories of an electron pair with zero total momentum $\vec{p}_1=-\vec{p}_2$ in separated graphene layers (separation $d=1.3$nm) with a hBN dielectric spacer ($\epsilon_s=3.9$). The constituent particles are in different colors, and the ticks on the axes correspond to steps of $20$nm.}
\label{trajectories_fig}
\end{figure}

In order to understand the dynamics of the electron pair in real space, it is instructive to analyze their semiclassical trajectories. Treating the momenta as classical variables, the time evolution of the two--particle system is then governed by Hamilton's equations:
\begin{align}
\frac{d\vec{p}_{i}}{dt}&=-\nabla_{\vec{x}_i}H,\nonumber\\
\frac{d\vec{x}_i}{dt}&=\nabla_{\vec{p}_i}H,\label{hamiltons}
\end{align}
where the subscript $i=1,2$ denotes the particle number.\\
\indent We will restrict our discussion to the case of indirect pairs in the subspace of states with configuration $\ket{2,\phi_{\vec{p}}}=\frac{1}{\sqrt{2}}[\ket{\uparrow\downarrow}+\ket{\downarrow\uparrow}]$. In this case the kinetic energy is isotropic and there is a trivial overlap of states with different momenta. The kinetic energy can be taken as the relevant eigenvalue of Eq. (\ref{single}) for each particle. Thus, we write:
\begin{align}
H=-\frac{p_1^2+p_2^2}{4m^*}+v_F(p_1-p_2)+\frac{V_0}{\sqrt{|\vec{x}_1-\vec{x}_2|^2+d^2}},\label{semi}
\end{align}
where $p_i=|\vec{p}_i|$ is the magnitude of the momentum for particle $i$, $\vec{x}_i=(x_i,y_i)$ is its position and $V_0=\frac{e^2}{\epsilon_s\epsilon_N^2}$ determines the strength of the potential. Then, Eqns. (\ref{hamiltons}) take the form:
\begin{align}
\frac{d\vec{p}_{1,2}}{dt}&=\pm\frac{V_0\vec{r}_{12}}{[|\vec{r}_{12}|^2+d^2]^{\frac{3}{2}}}\nonumber,\\
\frac{d\vec{x}_{1,2}}{dt}&=-\frac{\vec{p}_{1,2}}{2m^*}\pm v_F\frac{\vec{p}_{1,2}}{p_{1,2}}\label{systemeqns},
\end{align}
where $\vec{r}_{12}=\vec{x}_1-\vec{x}_2$. We solved Eqns. (\ref{systemeqns}) numerically using a $4^{\text{th}}$--order Runge--Kutta procedure, and a typical result is given in Fig. (\ref{trajectories_fig}).\\

\indent For a pair with vanishing total momentum, $\vec{p}_1=-\vec{p}_2\equiv\vec{p}$, Eqns. (\ref{systemeqns}) define the band velocities of the individual particles, which are given by
\begin{align}
\vec{v}_{1,2}=v_F\hat{\vec{p}}\mp\frac{\vec{p}}{2m^*},
\end{align}
where $\hat{\vec{p}}$ is a unit vector in the direction of $\vec{p}$. It immediately follows that the trajectories of the electrons can be represented in terms of a superposition of two motions: a fast center--of--mass motion, characterized by velocity $\vec{v}_{cm}=\frac{1}{2}\left(\vec{v}_1+\vec{v}_2\right)=v_F\hat{\vec{p}}$, and a much slower relative motion, with velocity  $\vec{v}_{r}=\vec{v}_2-\vec{v}_1=\vec{p}/m^*$. In the absence of interactions, and for small momenta, the relative velocity is sufficiently small that the pair behaves as if they were a single particle, moving with velocity $v_F$ in the direction of the momentum.\\
\indent If we now switch on the inter--particle interaction, the momenta of the particles become time--dependent, in accordance with Eqns. (\ref{systemeqns}). The sign of the interaction implies that whenever $\vec{p}_1$ is increasing, $\vec{p}_2$ must be decreasing, and vice versa. Due to the sign of the parabolic energy term, this will cause the particle with the smaller of the two velocities to speed up slightly, and the other particle to slow down, closing the distance between the two particles. Therefore, the finite relative velocity implies that a repulsive force between the two particles will increase the velocity of one particle while decreasing the velocity of the other, causing them each to change direction slightly in such a way that their seperation is almost constant. This causes the two particles to behave as if they were ``stuck together" despite the repulsive force. Note, however, that the momentum $\vec{p}$ is changed by the interaction. Hence, as the particles are orbiting around their mass center, the direction of $\vec{v}_{cm}$ is also changing, so that the average velocity of the pair over a long time is zero, as can be seen from Fig. (\ref{trajectories_fig}). The large value of $v_F$, compared to $\vec{v}_r$, results in non--propagating orbits of relatively large diameter $d\sim v_F/\omega$ where $\omega$ is the angular frequency of the orbit, which can be estimated as $\omega=E_0/\hbar$ ($d\sim 100$nm in Fig. (\ref{trajectories_fig})).\\
\section{Decay into the dispersing sector}
\indent So far we have considered only the non--dispersing sector. Coupling to the dispersing sector could lead to the decay of the metastable states found above, however these transitions are suppressed by momentum mismatch between the sectors. It is instructive to analyze the decay via Fermi's golden rule. Due to energy conservation the decay is only allowed into states with positive energy $E=2v_fp$, i.e. $\ket{3,\phi_{\vec{p}}}$ from Eq. (\ref{wf1}). The coupling between these two sectors occurs via trigonal--warping and potential energy terms, due to the non--trivial overlap between states with different momenta. The relevant matrix elements of the Hamiltonian are
$H_{1,3}=\frac{i}{\sqrt{2}}\tilde{V}_{\vec{p},\vec{p}'}\sin(\phi_{\vec{p}}-\phi_{\pvec{p}'})$ and $H_{2,3}=i\sqrt{2}\mu p^2\sin(3\phi_{\vec{p}})\delta_{\vec{p},\vec{p}'}$, where $H_{i,j}=\bra{i,\phi_{\pvec{p}'}}\widehat{H}\ket{j,\phi_{\vec{p}}}$.
\noindent The matrix element due to the interaction potential, $H_{1,3}$, vanishes by symmetry if $\psi_i(\vec{p})$ is an $s$--state. The kinetic energy term, $H_{2,3}$, conserves the momentum. Therefore the decay occurs when the initial ($p_i$) and final ($p_f$) momenta satisfy conservation laws: $E_0=2v_Fp_{i,f}$. This gives $p_i\ll p_0$, where $p_0=\sqrt{2\hbar m^{*}\omega}$ is the zero point momentum in the initial state. The smallness of $p_i$ results in a small matrix element which is proportional to $p^2$. The decay rate is given by Fermi's golden rule, $\Gamma=\frac{2\pi}{\hbar}|M_{if}|^2\rho_f$, where $M_{if}=\bra{\psi_f}H\ket{\psi_i}$ is the transition matrix element and $\rho_f=\frac{|E|}{8\pi\hbar^2v_F^2}$ is the density of final states\cite{wallace47} in the absence of spin--flipping and inter--valley scattering for particles with velocity $2v_F$. To calculate the kinetic contribution, we approximate the final state wavefunctions by plane waves, $\psi_f\sim\delta(\pvec{p}'-\vec{p}_f)$. The exact initial wavefunction for a direct pair state is not known, but for an order of magnitude estimate we shall employ the harmonic approximation as discussed above: $\psi_i(\vec{p})\sim \frac{1}{p_0}\exp(-\frac{p^2}{p_0^2})$. The transition rate due to $H_{2,3}$ is therefore $\Gamma=\frac{\mu^2p_f^4E_0}{\pi\hbar v_f^2p_0^2}\sim 10^{-10}\frac{E_0}{\hbar}$. This suggests only weak coupling to the continuum, justifying our consideration of the non--dispersing sector independently. Note that $H_{2,3}$ vanishes for indirect pairs, which do not decay by this mechanism.\\ 
\indent These small values imply that the lifetime of the pair is likely to be limited by other, non--universal mechanisms, such as impurity or electron--electron scattering. Also, one has to bear in mind that higher--order virtual transitions could lift the restriction $p_f=p_i\ll p_0$. The detailed analysis of this strongly depends on the properties of $\tilde{V}_{\vec{p},\pvec{p}'}$.\\
\section{Summary and conclusions}
\indent In this paper we have analyzed the problem of an isolated electron--electron bound state, however further work remains to be done in considering the many--body effects, the simplest example of which is interaction screening. Recall that in order to prevent Pauli blocking the system must be gated so that the Fermi energy exceeds the binding energy. This would introduce a non--zero density of states and metallic screening of the interaction. If the inter--particle separation is larger than the screening radius, the binding energy would be renormalized, but the bound state would not be destroyed: for a massive particle in an arbitrarily weak potential, at least one bound state exists in two dimensions\cite{landau77}. The states we considered here have positive energy and therefore do not represent energy minima, however once created they have long lifetimes. One way to create them is by coupling the aforementioned graphene structure to a superconductor. As the metastable states are akin to Cooper pairs, this would lead to a giant enhancement of the proximity effect, which has been observed in graphene recently \cite{ojeda09,du08,komatsu12}.\\
\indent In conclusion, we have studied the problem of interacting electron--electron pairs in hybrid graphene--dielectric--graphene structures. We have shown that, in the isotropic regime, the conduction--valence band asymmetry allows the formation of a new kind of Cooper--pair--like bound state in the sector spanned by eigenfunctions which are dispersionless in the conical approximation.\\
\section{Acknowledgements}
\indent The authors wish to thank V.I. Fal'ko and M.V. Berry for insightful discussions. A.V.S. is supported by EPSRC/HEFCE Grant No. EP/G036101/1.
\appendix
\section{Interlayer Coulomb interaction}
We consider the case of two interacting electrons, confined to separate graphene monolayers. We will begin by including the effects of screening in the random phase approximation (RPA). The ``screened'' interaction is given by
\begin{align}
V_q=\frac{v_qe^{-dq}}{(1+v_q\Pi_1)(1+v_q\Pi_2)-v_q^2\Pi_1\Pi_2e^{-2dq}}\label{coulomb},
\end{align}
(see Ref.~[\onlinecite{abergel2013}] for details), where $\Pi_1$ and $\Pi_2$ are the polarizabilities of layers $1$ and $2$ respectively, $v_q=\frac{2\pi e^2}{\epsilon_s q}$ is the bare Coulomb interaction and $d$ is the interlayer separation. We will show below that at the relevant distances the filling of the bands is not important, so that we use the approximation $k_f=0$ for both layers. Then $\Pi\equiv\Pi_{1,2}=\frac{Nq}{16v_f}$ (see Ref.~[\onlinecite{hwang07}] for details). Note that the Fermi energy can be tuned by application of a gate voltage in graphene, so we are free to assume the layers have equal carrier concentrations if we choose. The interaction then takes the form:
\begin{align}
V_q&=\frac{v_qe^{-dq}}{(1+v_q\Pi)^2-v_q^2\Pi^2e^{-2dq}}\\
&=\frac{v_qe^{-dq}}{1+2\frac{N\pi\alpha}{8}+(\frac{N\pi\alpha}{8})^2[1-e^{-2dq}]},
\end{align}
where $\alpha=\frac{e^2}{\epsilon_sv_f}$ is the Coulomb coupling constant.\\
\indent We note that we have assumed the Fermi energy is at the charge neutrality point ($E_f=0$) \textit{and} that the electrons are in opposite cones with opposite momenta. But if the valence band is full the phase space required to accommodate such a pair is occupied, and the state is therefore blocked due to the Pauli exclusion principle. A more detailed analysis of the many--body effects will be published elsewhere. To justify the choice of $k_f=0$ we note that the behaviour of $V(r)$ at distances $k_fr\lesssim1$ is determined by $q>k_f$ where the exact value of $k_f$ is not important. As we are most interested in the behaviour at small distances, $k_f=0$ is a reasonable first treatment of the two--body problem. This is a far more realistic approach than naively using the bare interaction, which results in unrealistically high binding energies.\\
\indent At small distances, $q\gg\frac{1}{d}$, the potential takes the form:
\begin{align}
V_q\sim\frac{v_q e^{-dq}}{1+2\frac{N\pi\alpha}{8}+\left(\frac{N\pi\alpha}{8}\right)^2}=\frac{2\pi e^2}{\epsilon_s\epsilon_N^2q}e^{-dq}.
\end{align}
The inverse Fourier transform is then a first approximation to the inter--layer potential in position space:
\begin{align}
V_r\sim\frac{e^2}{\epsilon_s\epsilon_N^2\sqrt{r^2+d^2}},
\end{align}
which is the form used to approximate the bound state energies in this work.\\
\indent At intermediate distances $d\ll r\ll\lambda_f$, however, the potential takes a slightly different form:
\begin{align}
V_q\sim\frac{v_q e^{-dq}}{1+2\frac{N\pi\alpha}{8}}=\frac{2\pi e^2}{\epsilon_s\epsilon_{N}q}e^{-dq},\label{intermediate}
\end{align}
where $\epsilon_N$ is defined in the same way as it was in the discussion above, but with $N=8$. We note the contribution of the two layers in Eq. (\ref{intermediate}), resulting in an extra ``which--layer" degree of freedom.
\section{Numerical approach}
\begin{figure}[h]
\includegraphics[scale=0.8]{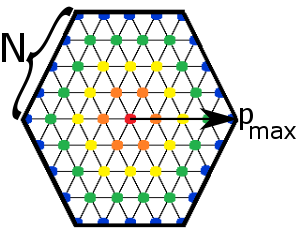}
\centering
\caption{(Color online) Schematic of the momentum--space grid for the example of $N=4$ shells. Shells are colored as follows: red ($0^{\text{th}}$ shell), orange ($1^{\text{st}}$ shell), yellow ($2^{\text{nd}}$ shell), and so on. The vector shown has magnitude $p_{\text{max}}$. All results in this paper were generated for $N=33$.}
\label{fig1}
\end{figure}

\indent The calculations utilized a triangular lattice of momentum--space sites to reflect the symmetry of graphene's honeycomb lattice. The grid is populated shell--by--shell (see Fig. (\ref{fig1})), so there are two parameters which can vary the results. The first is the number of shells, which we have taken to be $N=33$ for all results given in this paper (corresponding to approximately $3400$ grid points). There are a total of $3N(N+1)+1$ grid sites for $N$ shells, so increasing $N$ corresponds to increasing the total number of points, and thus reduces discretization errors. We did not obtain results for $N>33$ due to hardware limitations, but the energy levels were seen to vary negligibly above $N=20$. Secondly, one can vary the cut--off momentum $p_{\text{max}}$. Decreasing $p_{\text{max}}$ allows one to limit the phase space under consideration, hence increasing the density of sites. The results in the body of the present text were obtained for $p_{\text{max}}=\frac{\pi}{3a}$, thus covering the sites of primary interest in the low--energy physics.

In order to find the binding energies of the electron--electron pair, we begin with the discretized, momentum--space Schr\"odinger equation:
\begin{align}
H_{\text{eff}}(\vec{p})\psi_{\vec{p}}+S_0\sum_{\pvec{p}'}\tilde{V}_{\vec{p}-\pvec{p}'}A_{\vec{p},\pvec{p}'}\psi_{\pvec{p}'}=E\psi_{\vec{p}},\label{hamiltonianappendix}
\end{align}
\noindent which has been projected onto the subspace of non--dispersing states as explained above Eq. (3). We have defined $S_0$ to be the area of a cell in the momentum space grid and $A_{\vec{p},\pvec{p}'}$ is the matrix arising due to non--trivial overlap between non--dispersing states (see Eq. (4)). Note that Eq. (\ref{hamiltonianappendix}) can be re--written in the equivalent form:
\begin{align}
&\sum_{\pvec{p}'}\left[\begin{array}{cc}
-\frac{p^2}{2m^*}\delta_{\vec{p},\pvec{p}'}+S_0\tilde{V}_{\vec{p},\pvec{p}'}\cos(\phi_{\vec{p}-\pvec{p}'}) & 2\mu p^2\sin(3\phi_{\vec{p}})\delta_{\vec{p},\pvec{p}'}\nonumber \\
2\mu p^2\sin(3\phi_{\vec{p}})\delta_{\vec{p},\pvec{p}'}  & -\frac{p^2}{2m^*}\delta_{\vec{p},\pvec{p}'}+S_0\tilde{V}_{\vec{p},\pvec{p}'}
\end{array}\right]\psi_{\pvec{p}'}\\
&=E\psi_{\vec{p}}.\label{schrodinger}
\end{align}
\noindent The matrix in Eq. (\ref{schrodinger}) simply that of Eq. (4). We calculate the binding energies by populating a matrix with these blocks or, equivalently, by constructing a system of equations of the form (\ref{schrodinger}), each of which is designated a unique momentum $\vec{p}$.\\

\end{document}